\documentclass[twocolumn,showpacs,preprintnumbers,nofootinbib,amsmath,amssymb]{revtex4}
\usepackage{amsmath,amssymb,amsfonts}
\usepackage{graphics}
\usepackage{epsfig}

\newcommand\Ga{\Gamma}

\newcommand\lam{\lambda}

\newcommand\rmi{{\rm i}}

\begin{document}
\title{In-medium bound states and  pairing gap}
\author{O.A. Rubtsova}
\email{rubtsova@nucl-th.sinp.msu.ru}
\author{V.I. Kukulin}
\email{kukulin@nucl-th.sinp.msu.ru}
\author{V.N. Pomerantsev}
\email{pomeran@nucl-th.sinp.msu.ru} \affiliation{Skobeltsyn
Institute of
 Nuclear Physics, Moscow State University, Leninskie gory, Moscow,
119991, Russia}
\author{H. M{\"u}ther}
\email{herbert.muether@uni-tuebingen.de}
 \affiliation{Institute for Theoretical Physics, University of T\"ubingen, Auf der
Morgenstelle 14, D-72076 T\"ubingen, Germany}
\begin{abstract}

The propagator of two nucleons in infinite nuclear matter is evaluated by a diagonalization of the $pphh$
RPA Hamiltonian. This effective Hamiltonian is non-Hermitian and, for specific density domains and partial
waves, yields pairs of complex conjugated eigenvalues representing in-medium bound states of two nucleons.
The occurrence of these complex poles in the two-particle Greens function is tightly related to the well
known BCS pairing approach. It is demonstrated that these complex eigenvalues and the corresponding bound
state wavefunctions contain all information about the BCS gap function. This is illustrated by calculations
for $^1S_0$ and $^3PF_2$ pairing gaps in neutron matter which essentially coincide with the corresponding
gap functions extracted from conventional solutions of the gap equation.
Differences between the bound states in the
conventional BCS approach and the $pphh$ RPA are arising in  the case of $^3SD_1$
channel in symmetric nuclear matter at low densities. These differences are discussed in the context of
transition from BEC for quasi-deuterons to the formation of  BCS pairing.

\end{abstract}
 \maketitle

 \section{Introduction} The pairing of fermions in
Fermi liquids has been studied in detail many decades ago and became
an important part of fundamental science and a corner stone of the theories on
superfluidity and superconductivity \cite{Abrikosov,
Schrieffer}.
In the conventional way in BCS method~\cite{Schrieffer} or Bogolyubov's
approach one derives and solves the non-linear gap equation
for the gap function $\Delta(k)$ which describes the deviation of the single particle
energies from continuous spectrum near the Fermi-surface. This procedure has
successfully been used in solid state physics and later has been applied also for
studying superfluidity in nuclear matter \cite{baldo90,elgor90,kuckei}. The values for the gap in the
$^1S_0$ channel for neutron matter extracted from such
calculations turned out to be in a reasonable
agreement with empirical data for neutron-neutron and proton-proton pairing in finite nuclei. In neutron matter
at densities above the saturation density of symmetric nuclear matter the pairing of neutrons in the $^3PF_2$
channel occurs also and may become relevant for understanding the cooling of neutron stars \cite{page,wcgho,Ram2}.

The interaction between proton and neutron in the triplet $^3SD_1$ channel, however, is more attractive then
the interaction between two neutrons in $1S_0$ channel and leads to the vacuum bound state of two
nucleons in the deuteron channel $^3SD_1$. Therefore one would expect even stronger pairing effects from
proton-neutron pairing in isospin symmetric nuclear systems. Indeed, BCS calculations for symmetric nuclear
matter lead to a sizable gap of around 10 MeV at the empirical saturation density
\cite{alm90,vonderf,Baldo_deu,wim05}. The empirical data for finite nuclei with number of protons close to
the number of neutrons, however, do not show any indications of strong proton neutron pairing, which would
correspond to a pairing gap as large as 10 MeV.

Therefore attempts have been made to embed the BCS approach into more general scheme for evaluating
self-consistent Greens's functions (SCGF). Usually this is done in the framework of a generalized mean field
approach within the well known Nambu--Gorkov formalism with an explicit BCS parameterization for single
particle energies and the respective Green's functions. More generally \cite{Dickhoff} one tries to solve
SCGF equations in the $T$-matrix or ladder approximation with a self-consistent evaluation for
single-particle and two-particle Green functions. Employing realistic models for the nucleon-nucleon (NN)
interactions, i.e. NN potentials which fit the NN scattering data such as the Argonne V18 \cite{v18} or
CD-Bonn potential \cite{cdbonn}, one obtains significant deviations from the mean-field results. The
strong short-range and tensor components of such realistic NN interactions lead to non-negligible depletions
of the occupation of states with momenta below the Fermi momentum $k_F$ and corresponding occupations of
states with momenta larger $k_F$.

The central equation to be solved with such SCGF calculations is scattering or $T$-matrix equations for a
pair of interacting nucleons in nuclear matter.  The resulting $T$-matrix is used to define the nucleon
self-energy, which is needed to evaluate the single-particle (sp) Green's function. Since the information on
energy- and momentum-distribution of the single-particle strength, which is contained in the sp Green's
function, is required to set up $T$-matrix equation, the evaluation of the sp Green's function and the
solution of the corresponding $T$-matrix equation have to be done in a self-consistent manner. The solution of the
$T$-matrix integral equation is complicated due to the so-called pairing instabilities
which are related to the occurrence of quasi-bound two-nucleon states in the nuclear medium \cite{frick05}.

On the other hand,  we have recently developed  a formalism in which the two-particle Green's function is
evaluated in terms of discrete eigenvalues and eigenfunctions of a two-particle Hamiltonian  \cite{NM1}.
This approach allows to treat two-particle continuum and two-particle bound states in vacuum and also in medium
on the same footing. It has been applied to evaluate the NN scattering phase-shifts as well as the solution
of Bethe--Goldstone equation, the nuclear $G$-matrix, in an efficient way. In the present paper we will
generalize this approach to include not only the particle-particle states ($pp$), as it has been done for
Bethe--Goldstone equation, but also the hole-hole states ($hh$) in evaluating the in-medium $T$-matrix. This
approach leads to the diagonalisation of an effective Hamiltonian, which corresponds to the $pphh$
RPA Hamiltonian.

The eigenvalues of  this  Hamiltonian become complex in the region of pairing instabilities. This implies
that the occurrence of such instabilities in the framework of the SCGF is under control. In the mean-field
limit for the single-particle Green's function the RPA equation for two nucleons with c.m. momentum equal to
zero correspond to the BCS approach. Here it is worth to mention the novel stabilization
technique\cite{Ram1}. In this technique one determines the gap function at the Fermi surface
from the imaginary part of the complex eigenvalues of $pphh$ RPA.

In the present paper we will demonstrate that   not only the complex eigenvalues of the effective
Hamiltonian can be used in finding the gap in the vicinity of Fermi surface but the corresponding
eigenfunctions of these eigenvalues are also directly related to the momentum dependence of the gap function for
all momenta. This yields a new method for finding the gap function which is efficient in particular for
pairing in coupled channels such as $^3PF_2$ in the case of neutron-neutron pairing or $^3SD_1$ for
proton-neutron pairing.

 For the sake of simplicity we restrict our study in the present work  to the case of zero temperature and
use the independent particle limit for the single-particle Green's function. The present approach  can be
also extended to non-zero temperature  and employment of dressed single-particle Green' s function in the
SCGF approach \cite{NM2}.

In the next section we will show how the complex eigenvalues and eigenfunctions of
the $pphh$ RPA equation are related to the gap function derived from the BCS approach. Applications for this
new method to solve the BCS equation in the case of neutron matter will be discussed in section III and
section IV shows results for proton-neutron pairing in the quasi-deuteron channel of symmetric nuclear
matter. Conclusions are presented in the final section V.

\section{Effective two-body Hamiltonian.}
\subsection{Equation for the $T$-matrix}
 Consider the equation for the in-medium $T$-matrix in
which hole-hole ($hh$) degrees of freedom are also included:
\begin{equation}
\label{T-matrix} T(E)=V+VG_{II}^0(E)T(E),
\end{equation}
where $V$ is a bare interaction and $G_{II}^0$ is the non-interacting two-body
$pphh$-propagator
\begin{equation}
\label{G0} G^0_{II}(E)=\int d{\bf k}_1 d{\bf k}_2|{\bf k}_1,{\bf
k}_2\rangle G_{II}^0(k_1,k_2;E) \langle {\bf k}_1,{\bf k}_2|\,.
\end{equation}
If we consider the mean-field approximation
the kernel takes the form \cite{Dickhoff}:
\begin{eqnarray}
G_{II}^0(k_1,k_2;E)=\frac{\theta(k_1-k_F)\theta(k_2-k_F)}{E+\rmi
0-(e_{k_1}+e_{k_2})}-\nonumber\\
-\frac{\theta(k_F-k_1)\theta(k_F-k_2)}{E-\rmi 0-(e_{k_1}+e_{k_2})}.
\end{eqnarray}
Here $k_1$ and $k_2$ are single particle momenta, $e_k$ are single
particle energies, $k_F$ is the Fermi-momentum and all the states
are antisymmetrized.

 The operator $G_{II}^0$ can
 formally be written as a generalized resolvent for the
free Hamiltonian which includes both $pp$- and $hh$- continuum
contributions:
\begin{eqnarray}
H_0=\int_{{k_1,k_2}\leq k_F}d{\bf k}_1 d{\bf k}_2|{\bf k}_1,{\bf
k}_2\rangle[e_{k_1}+e_{k_2}] \langle {\bf k}_1,{\bf k}_2|\nonumber\\
+\int_{k_1,k_2>k_F}d{\bf k}_1 d{\bf k}_2|{\bf k}_1,{\bf
k}_2\rangle[e_{k_1}+e_{k_2}] \langle {\bf k}_1,{\bf k}_2|.
\end{eqnarray}
 So that, one can express the $pphh$-propagator through $H_0$ Hamiltonian:
\begin{equation}
\label{g0J} G_{II}^0(E)=[EJ+{\rmi}0-JH_0]^{-1},
\end{equation}
where $J$ is the operator with the following representation in
momentum space:
\begin{equation}
J(k_1,k_2)=\theta(k_1-k_F)\theta(k_2-k_F)-\theta(k_F-k_1)\theta(k_F-k_2),
\end{equation}

 Further, the two-body $T$-matrix (\ref{T-matrix}) can be rewritten in a
form
\begin{equation}
T(E)=V+VG_{II}(E)V,
\end{equation}
where we introduced the operator $G_{II}$  which should satisfy the
integral equation: \begin{equation}
G_{II}=G_{II}^0+G_{II}^0VG_{II}=[G_{II}^0(E)^{-1}-V]^{-1}.\end{equation}
By using the expression  (\ref{g0J}) one gets the following form of
the above operator:
\begin{equation}
\label{gJ} G_{II}(E)=[EJ+{\rmi}0-JH]^{-1},
\end{equation}
where
\begin{equation}
 \label{H}
H=H_0+JV\,. \end{equation}
 By comparing eqs.~(\ref{g0J}) and (\ref{gJ}) one concludes that the Hamiltonian
  $H$  can be
considered as an effective two-body Hamiltonian describing
interaction of particles and holes.  This effective Hamiltonian corresponds to the Hamiltonian of the $pphh$ RPA.

  The accurate formalism for a practical
treatment of this type operator and the evaluation of the $T$-matrix will be published elsewhere
\cite{NM2}. In the present paper, we focus  on the study of the
 bound states of this effective Hamiltonian in a case of
zero center of mass momentum $K=0$ and concentrate on the occurrence of pairing
phenomena.

 \subsection{In-medium bound states and the gap equation}
 Because the total Hamiltonian (\ref{H})  is non-Hermitian but real
it may have pairs of complex conjugated eigenfunctions
$|\psi\rangle$ and $|\psi^*\rangle$ and respective complex
conjugated eigenenergies $E_b$ and $E_b^*$.

We emphasize that the bound state $|\psi\rangle$ of $H$ is at the
same time the eigenfunction of the homogeneous equation
corresponding to eq.~(\ref{T-matrix}), i.e.:
\begin{equation}
\label{Sch_eq} H|\psi\rangle=E_b|\psi\rangle \Longleftrightarrow
G_{II}^0(E_b)V|\psi\rangle=\eta(E_b)|\psi\rangle,
\end{equation}
with unit  eigenvalue $\eta(E_b)=1$. The latter equation has been
studied in the stabilization approach \cite{Ram1,Ram2}, where
the positions of the eigenvalues $E_b=E_0+{\rmi}\Ga_0$  in a complex
energy plane have been examined. The above approach uses
the fact that the real part of the eigenvalue $E_0\approx2e_F$
(where $e_F$ is the Fermi-energy) while the imaginary part $\Ga_0$
coincides with the pairing gap $\Delta(k_F)$ at the Fermi-momentum.

We will demonstrate below that the bound state wave functions are
also important objects as they are related directly to the momentum
dependence of the pairing gap $\Delta(k)$.
 The Schroedinger equation on the  left from arrow in eq.~(\ref{Sch_eq})
   is reduced to the following  system in the representation of the relative
  momentum $k$:
\begin{equation}
\label{psi_eq} J_k(2e_k-E_0-\rmi \Ga_0)\psi(k)=-\int dk' {k'}^2
V(k,k') \psi(k'),
\end{equation}
where factor $J_k=\theta(k-k_F)-\theta(k_F-k)$ is equal to 1 or $-1$
for $pp$ and $hh$ parts of the continuum respectively and the integral is taken over
all intermediate momentum states. The solution of the
eq.~(\ref{psi_eq}) can be rewritten in a form:
\begin{equation}
\label{Gak} \psi(k)=\frac{f(k)}{J_k(2e_k-E_0-\rmi \Ga_0)}.
\end{equation}
The equation (\ref{psi_eq}) has a similar form to  the equation for
the gap wave function (or anomalous density) $\chi(k)$ in the BCS
approach:
\begin{eqnarray}
\label{Dek} \chi(k)2E_k=-\int dk' {k'}^2
{V(k,k')\chi(k')},\\
\nonumber \chi(k)=\frac{\Delta(k)}{2E_k},\quad
E_k=\sqrt{(e_{k}-e_F)^2+\Delta^2(k)}
\end{eqnarray}
where $\Delta(k)$ is the gap function.

We note the interesting fact (which  can be easily  proven
 for a case when the bare interaction is given by a separable potential
$V=\lam|\varphi\rangle\langle \varphi|$) that solutions of the
eqs. (\ref{psi_eq}) and (\ref{Dek}) are interrelated to each other up
to energy terms, so that one has an approximate formula:
\begin{equation} \label{fd}|f(k)|\approx A|\Delta(k)|, \end{equation}
where $A$ is some normalization constant. Here we use absolute
value of function because $f(k)$ is complex.

 For  the case
of a realistic bare interaction $V$ we found that the relation
(\ref{fd}) is valid in a very good approximation. It  is likely correct to the extent  that one uses
the approximation of the fixed gap in the left hand side of the
eq.~(\ref{Dek}), i.e. $E_k=\sqrt{(e_k-e_F)^2+\Delta^2(k_F)}$.

On the basis of the eq.~(\ref{fd}), we may suggest that the absolute
value of the bound state wave function  takes the following form:
\begin{equation}
\label{z_bound} |\psi(k)|\approx A\phi(k),
\end{equation}
where $A$ is a normalization factor and $\phi(k)$ is a
characteristic function defined by:
\begin{equation}
\label{norm} \phi(k)= \frac{|\Delta(k)|}{\sqrt{(2e_k-E_{\rm
0})^2+\Ga_{0}^2}},\quad \int_{0}^\infty
|\phi(k)|^2k^2dk=\frac{1}{A^2}.
\end{equation}
Below we will check  this relation numerically and will use it to
determine the gap functions in neutron and  symmetric nuclear matter.
\section{Neutron matter case}
For practical calculations we have used the discrete stationary
wave-packet basis (SWP) \cite{NM1}  which corresponds to a
discretization of the relative momentum $k$, so that, the
eigenfunctions of $H$ can be found from a diagonalisation procedure
for its matrix in the SWP basis. All calculations in the following sections have been performed using the CD Bonn potential \cite{cdbonn}
and considering the kinetic energy for the single-particle energies $e_k$.

\subsection{Spin-singlet channel}
\begin{figure}[h]
\epsfig{file=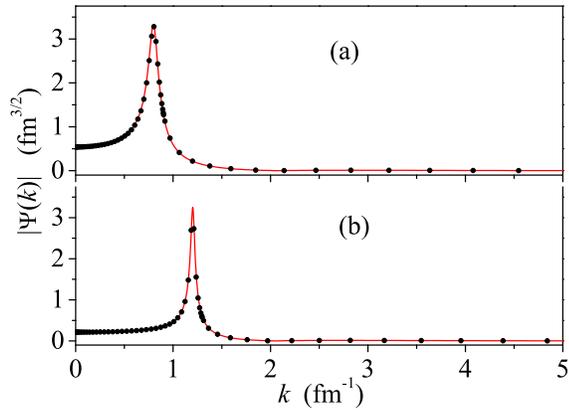,width=\columnwidth} \caption{(Color online) The
bound-state wave function of the Hamiltonian $H$ (solid curve) in comparison
with characteristic function $A\phi(k)$ (filled circles) at $k_F$
=0.8 (a) and 1.2 fm$^{-1}$ (b).  \label{fun_singlet}}
\end{figure}

 In Fig.~\ref{fun_singlet}
 comparisons of the functions $\psi(k)$ and $A\phi(k)$ are presented for
the spin-singlet $^1S_0$ channel in neutron matter at $k_F$=0.8 and
1.2 fm$^{-1}$. Both functions are normalized to unity in momentum
space. Here the function $\phi(k)$ is defined from the pairing gap
$\Delta(k)$ calculated from solution of the BCS gap equation in a conventional way while the function
$\psi(k)$ is found from a direct diagonalisation procedure for the
total Hamiltonian $H$ matrix in the stationary wave-packet
basis \footnote{The details of the procedure will be published
elsewhere \cite{NM2}.} \cite{NM1}. It is evident from
Fig.~\ref{fun_singlet} that
 both functions are almost indistinguishable.

If one considers furthermore in dependence on the stabilization approach
\cite{Ram1,Ram2} that $\Ga_0$ defines the pairing gap at the
Fermi-momentum, it
is possible to extract the momentum dependence of the gap $\Delta(k)$ just
from our bound state wave function:
\begin{equation}
\label{gap_d}
|\Delta(k)|\approx\frac{|\psi(k)|}{|\psi(k_F)|}\sqrt{(2e_k-E_0)^2+\Ga_0^2}.
\end{equation}
\begin{figure}[h!]
\epsfig{file=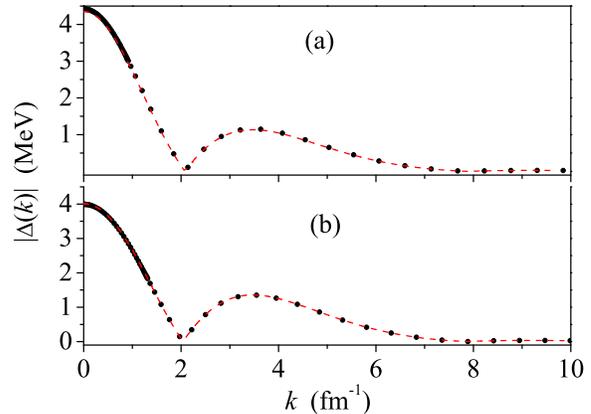,width=\columnwidth} \caption{(Color online) The absolute
value of the gap $\Delta(k)$ found from the bound state wave
function (dashed curves) in comparison with direct solutions of the
gap-equation (filled circles) for the $^1S_0$ channel at $k_F$ =0.8
fm$^{-1}$ (a) and 1.2 fm$^{-1}$ (b). \label{gaps_singlet}}
\end{figure}
In Fig.~\ref{gaps_singlet} the gaps (the absolute values) for the
$^1S_0$ channel found from the formula (\ref{gap_d}) are presented in comparison
with direct solutions of the non-linear integral gap equation for
$k_F=0.8$ and 1.2 fm$^{-1}$.

The agreement is excellent. Therefore, we may conclude that the bound
state wave function of the total Hamiltonian contains all the
information about the pairing gap function.

\subsection{Spin-triplet channel}
In a case of the coupled channel $^3PF_2$, the bound state wave
function can be represented as a sum of the partial wave
contributions:
\begin{equation}
|\psi(k)|^2=|\psi^{l=1}(k)|^2+|\psi^{l=3}(k)|^2.
\end{equation}
However, the general structure is the same as in the singlet channel
case, i.e.
\begin{equation}
|\psi(k)|^2\approx A^2 \phi^2(k),\quad
\phi^2(k)=\frac{\Delta_{l=1}^2(k)+\Delta_{l=3}^2(k)}{(2e_k-E_0)^2+\Ga_0^2},
\end{equation}
which is consistent with the relation for the two-channel total gap
$\Delta^2(k)=\Delta_{l=1}^2(k)+\Delta_{l=3}^2(k)$.

\begin{figure}[h!]
\epsfig{file=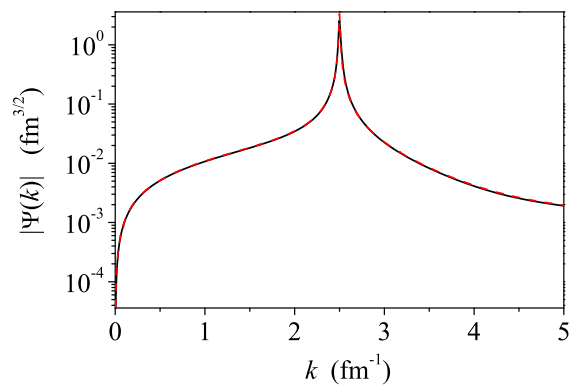,width=\columnwidth} \caption{(Color online) The
bound-state wave function of the RPA Hamiltonian $H$ (solid curve) in comparison
with characteristic function $A\phi(k)$ (dashed curve) for the the
 $^3PF_2$ channel in neutron matter at $k_F=2.5$ fm$^{-1}$.  \label{fun_SP}}
\end{figure}

 We treat the coupled-channel case in
Fig.~\ref{fun_SP} where the bound state wave functions for the
$^3PF_2$ channel are displayed in comparison with characteristic
wave functions $\phi(k)$ (normalized to unity). Here the pairing gap
found from the solution of the coupled-channel gap-equation is used
to find the function $\phi(k)$. The agreement is again perfect.

\begin{figure}[h!]
\epsfig{file=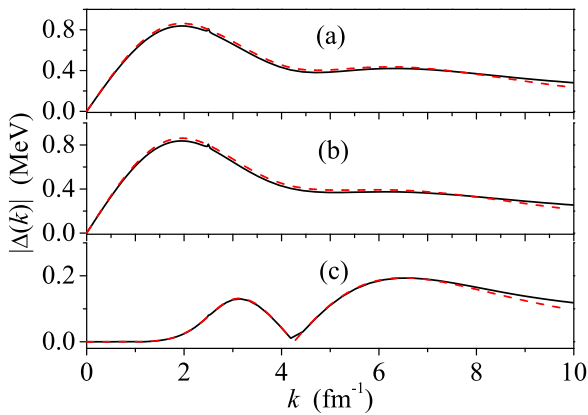,width=\columnwidth} \caption{(Color online) The absolute
value of the gap $\Delta(k)$ found from the bound state wave
function (dashed curves) in comparison with solutions of the
gap-equation (solid curves) for the $^3PF_2$ channel at $k_F$ =2.5
fm$^{-1}$ (a) and the partial gaps for the partial $P$ (b) and $F$
(c) waves.
 \label{gap_pf}}.
\end{figure}

A direct comparison of the gaps for $^3PF_2$ channel extracted from
the bound-state wave functions and those found from a direct
solution of the gap integral equation is shown in Fig.~\ref{gap_pf}.

Thus, it appears  that the eigenfunctions of the total Hamiltonian
corresponding to the complex-valued bound states near the Fermi
surface contain all the required information about the pairing gap.
This result demonstrates clearly that a solution of the nonlinear gap
equation can be replaced (at least in a case of neutron matter) with
much simpler solving for the effective eigenvalue problem even in a
case of coupled channels.

\section{The $^3SD_1$ channel in symmetric nuclear matter}
To make the picture more complete, we also consider the case of
$^3SD_1$ channel in symmetric nuclear matter. The main difference
as compared to the cases discussed above for neutron matter is
the more attractive interaction in this channel, which leads to the bound state
of the deuteron in the limit of zero density.

%\subsection{Problems at low densities}

\begin{figure}
\epsfig{file=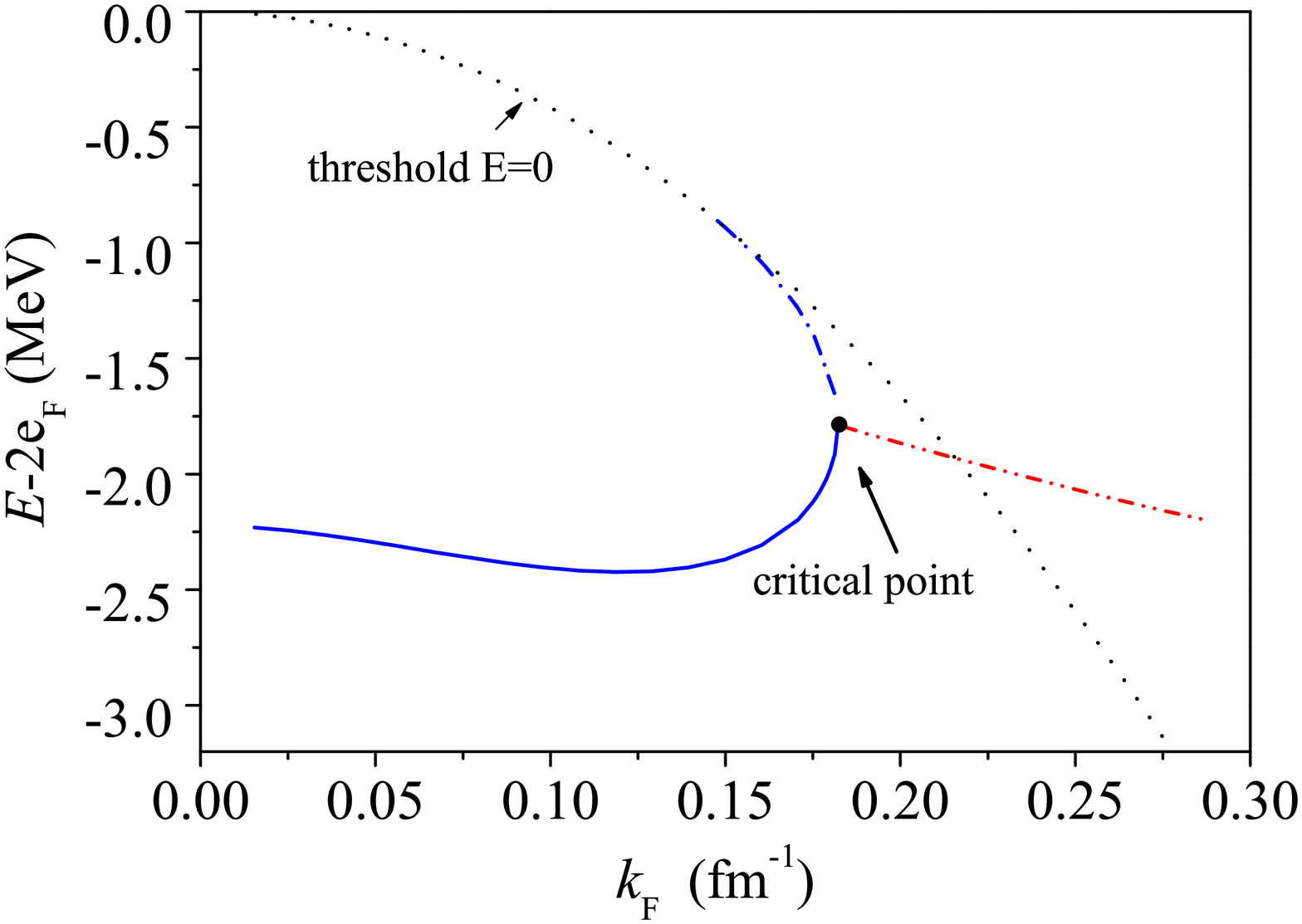,width=\columnwidth}
 \caption{(Color online) Behavior of the real-valued bound-state energies (blue curves) and the real part of
  the complex-valued bound-state energies  (red curve)
   for small $k_F$ values
   in the spin-triplet $NN$
channel at $K=0$.\label{fig_bound}}
\end{figure}

The evolution of the bound state energy of $H$ with increasing Fermi momentum $k_F$
is displayed  in Fig.~\ref{fig_bound}. The energy scale in this
figure is chosen in such a way that the continuum for two hole states
is above the dotted line denoted with "threshold E=0".

At zero $k_F$  the Hamiltonian $H$ has the same bound state as the
the $NN$ Hamiltonian in free space. It corresponds to the energy of
the deuteron at around -2.23 MeV. With increasing $k_F$, the bound
state energy $E_b$ of $H$ still remains below the $hh$ threshold and
thus is  real (the blue solid curve in Fig.~\ref{fig_bound}).
Moreover, at some density (at $k_F\sim 0.14$ fm$^{-1}$), {\em the
second} real bound-state arises under the threshold $E=0$ (the blue
dash-dotted curve). Then, at another critical density $k_F^C\approx
0.182$ fm$^{-1}$ these two bound states merge to one point (the
black filled circle) and for the higher $k_F$ they are transformed
into the pair of complex conjugated bound states $E_{b}$ and $E_b^*$
(the real part of these eigenvalues is denoted by the red
dash-dot-dotted curve).  Thus, at very low density our total
Hamiltonian treatment leads to a picture that differs from the
conventional BCS approach. The main difference is related to the
fact that for Fermi momenta below a critical value $k_F^C$ bound
states occur with real energies below the threshold of the $hh$
continuum.

This may lead to the conclusion  that at densities with Fermi
momenta below $k_F^C$ the formation of bound quasi-deuterons is
 energetically favorable as compared to the formation of BCS Cooper pairs.
 At these low densities the quasi-deuterons may form a Bose--Einstein Condensate (BEC) of
deuterons with zero total momentum \cite{Matsuo,armen} and the critical Fermi momentum $k_F^C$
would be interpreted to describe the phase transition from BEC to BCS. It should be noted,
 however, that our estimate for this BEC--BCS transition is not very realistic in the sense
that we ignore the Coulomb interaction between protons, the contributions of electrons and
 the formation of isospin asymmetric nuclear matter as well as the possibility to form nuclear cluster larger than the deuteron as e.g. $\alpha$-cluster.

\begin{figure}
\epsfig{file=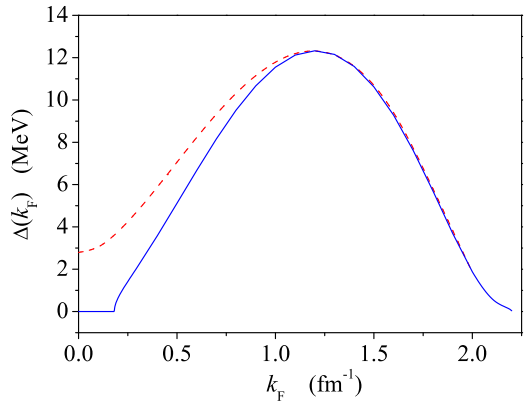,width=\columnwidth}
 \caption{(Color online) The gap value $\Delta(k_F)$ found from the solution of the gap equation
 (dashed curve) and imaginary part of the eigenvalue $E_b$ for the $^3SD_1$ channel
 in symmetric nuclear matter.\label{gap_sd_low}}.
\end{figure}

 In Fig.~{\ref{gap_sd_low}} we compare the pairing gaps at $k_F$ found from the solution of the gap equation with imaginary
parts of the total Hamiltonian eigenvalues. For the region $k_F < k_F^C$ at which $E_b$ is real, the latter value is stated as
zero. One can see from this figure that the pairing gap $\Delta(k_F)$ derived from the imaginary part of the $pphh$ RPA
eigenstate is below the pairing gap obtained from a conventional solution of the BCS equation also at densities above $k_F^C$.
This is probably related to the fact that also for these densities the real part of the lowest eigenvalue $E_0$ is
significantly below the $2\varepsilon_F$. This may indicate that the BCS approximation  is not appropriate to describe the
strong $NN$ correlations between protons and neutrons at these densities. This would also explain why one does not observe any
features of proton-neutron BCS pairing in finite nuclei.

\begin{figure}
\epsfig{file=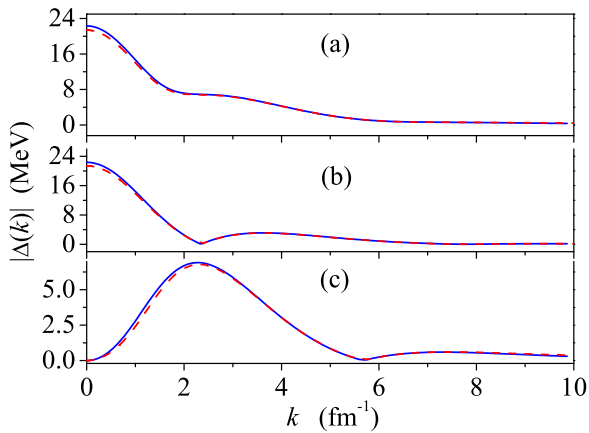,width=\columnwidth}
 \caption{(Color online) The total pairing gap $\Delta(k)$ (a) and the partial $S$ (b) and $D$ (c) gaps
   found from the solution of the gap equation
 (dashed curve) and from the bound state wave function for the $^3SD_1$ channel
 in symmetric nuclear matter at $k_F$=1.2 fm$^{-1}$.\label{gap_sd_12}}.
\end{figure}

The Fig.~{\ref{gap_sd_low}} also reflects  the fact that at $k_F$
around 1.2~fm$^{-1}$, which is not far from the saturation density,
the results for $\Delta(k_F)$ coincide.
  The same is also true for the momentum dependence of the pairing
gap $\Delta (k)$ as can be seen from  Fig.~{\ref{gap_sd_12}}.

\section{Summary}
In the paper we have developed a simple technique which allows to
evaluate the function of BCS pairing gap $\Delta(k)$ in terms of the
complex eigenvalue and corresponding eigenfunctions of the $pphh$
RPA Hamiltonian. With realistic $NN$ interactions this approach
provides results for the gap function in neutron matter which are in
a very good agreement with those derived from a conventional
solution of the BCS equation.

The study of the bound states of the $pphh$ RPA Hamiltonian yields
results different from the BCS approach if one considers the strong
two-nucleon correlations between protons and neutrons indicating a
transition from a BEC of quasi-deuterons to the formation of
corresponding BCS pairs.

The connection between the BCS approach and the $pphh$ RPA
Hamiltonian established here leads to a generalization of the BCS
approach into a  treatment of two-particle correlations within the
scheme of self-consistent calculations for one- and two-particle
Green's function (SCGF).

In this short paper we present results for the approach which treats
the single-particle Green's function in the mean-field
approximation. However the discussing model contains important
features of a general treatment.
   Thus, the approach can be generalized to a realistic case in a straightforward manner.
 The corresponding investigations are in preparation.

{\bf Acknowledgments.} The authors appreciate  the financial
support from the DFG grant MU 705/10-1, the joint DFG--RFBR grant
16-52-12005 and the RFBR grant 16-02-00049.

\end{document}